\documentclass[iop,apj,twocolumn]{emulateapj} \usepackage{times}

\usepackage{graphicx}
\begin{document}


\newcommand{\etal}{et al.}
\newcommand{\fermi}{\emph{Fermi}}
\newcommand{\fermilat}{\emph{Fermi}--LAT}
\newcommand{\gray}{$\gamma$-ray}
\newcommand{\imos}[1]{({\bf IMOS: #1})}
\newcommand{\change}[1]{{\bf #1}}

\slugcomment{Accepted by The Astrophysical Journal}

\shorttitle{\fermilat{} observations of the quiescent sun}
\shortauthors{Abdo et al.}

\title{\fermilat{} Observations of Two Gamma-Ray Emission Components from the Quiescent Sun
}
\author{
A.~A.~Abdo\altaffilmark{1}, 
M.~Ackermann\altaffilmark{2}, 
M.~Ajello\altaffilmark{2}, 
L.~Baldini\altaffilmark{3}, 
J.~Ballet\altaffilmark{4}, 
G.~Barbiellini\altaffilmark{5,6}, 
D.~Bastieri\altaffilmark{7,8}, 
K.~Bechtol\altaffilmark{2}, 
R.~Bellazzini\altaffilmark{3}, 
B.~Berenji\altaffilmark{2}
% ...
}
\altaffiltext{1}{Center for Earth Observing and Space Research, College of Science, George Mason University, Fairfax, VA 22030, resident at Naval Research Laboratory, Washington, DC 20375}
\altaffiltext{2}{W. W. Hansen Experimental Physics Laboratory, Kavli Institute for Particle Astrophysics and Cosmology, Department of Physics and SLAC National Accelerator Laboratory, Stanford University, Stanford, CA 94305, USA}
\altaffiltext{3}{Istituto Nazionale di Fisica Nucleare, Sezione di Pisa, I-56127 Pisa, Italy}
\altaffiltext{4}{Laboratoire AIM, CEA-IRFU/CNRS/Universit\'e Paris Diderot, Service d'Astrophysique, CEA Saclay, 91191 Gif sur Yvette, France}
\altaffiltext{5}{Istituto Nazionale di Fisica Nucleare, Sezione di Trieste, I-34127 Trieste, Italy}
\altaffiltext{6}{Dipartimento di Fisica, Universit\`a di Trieste, I-34127 Trieste, Italy}
\altaffiltext{7}{Istituto Nazionale di Fisica Nucleare, Sezione di Padova, I-35131 Padova, Italy}
\altaffiltext{8}{Dipartimento di Fisica ``G. Galilei", Universit\`a di Padova, I-35131 Padova, Italy}
% ...

\begin{abstract}

We report the detection of high-energy \gray{s} from the quiescent Sun with the Large Area Telescope (LAT)
on board the \emph{Fermi Gamma-Ray Space Telescope} (\fermi) during the first 18 months of the mission.
These observations correspond to the recent period 
of low solar activity when the emission induced by cosmic rays is brightest.
For the first time, the high statistical significance of the observations 
allows clear separation of the two components:
the point-like emission from the solar disk due to cosmic ray 
cascades in the solar atmosphere, and extended 
emission from the inverse Compton scattering of cosmic ray electrons on solar photons in the heliosphere. 
The observed integral flux ($\ge$100 MeV) from the solar disk is $(4.6 \pm 0.2[{\rm statistical\ error}]^{+1.0}_{-0.8}[{\rm systematic\ error}]) \times10^{-7}$ cm$^{-2}$ s$^{-1}$, 
which is $\sim$7 times higher than predicted by the ``nominal" model of \citet{Seckel1991}.
In contrast, the observed integral flux ($\ge$100 MeV) of the \emph{extended} emission
from a region of $20^{\circ}$ radius centered on the Sun, but excluding the disk itself,
$(6.8\pm0.7[{\rm stat.}]_{-0.4}^{+0.5}[{\rm syst.}]) \times 10^{-7}$ cm$^{-2}$ s$^{-1}$, along with the observed spectrum and the angular profile, 
are in good agreement with the theoretical predictions for the inverse 
Compton emission.

\end{abstract}


\maketitle

\section{Introduction} \label{sec:introduction}

The Sun is a well-known 
source of X-rays  and \gray{s} during solar flares \citep{Peterson1959,Chupp1973,Kanbach1993}, 
which are high-energy phenomena associated with the flare-accelerated
particle interactions in the solar atmosphere. Quiescent solar \gray{} emission from hadronic cosmic-ray (CR) interactions with the solar
atmosphere and photosphere was first mentioned by \citet{Dolan1965}. 
\citet{Peterson1966} estimated its flux based on measurements of terrestrial emission, and \citet{Hudson} 
suggested it to be detectable ($\sim$$10^{-7}$ cm$^{-2}$ s$^{-1}$ above 100 MeV) by the EGRET experiment on board 
the \emph{Compton Gamma-Ray Observatory} (\emph{CGRO}). 
The first, and so far the only, detailed theoretical study of \gray{} emission from interactions of CR protons in the
solar atmosphere was published by 
\citet{Seckel1991}. The integral flux above 100 MeV was predicted to be $F(\ge100\ {\rm MeV}) \sim (0.22-0.65)\times10^{-7}$ cm$^{-2}$ s$^{-1}$ for their ``nominal'' model.
However, attempts to observe such emission with EGRET (1991--1995) yielded only an upper limit of $2.0 \times 10^{-7}$ cm$^{-2}$ s$^{-1}$ 
above 100 MeV at $95\%$ confidence level \citep{Thompson}.

The existence of an additional, spatially extended component of the solar emission due to the inverse Compton (IC) scattering of CR electrons off solar photons was not 
realized until recently \citep{Moskalenko2006,Orlando2007}. While the IC
emission is brightest in the region within a few degrees of the Sun (hereafter we refer the angle relative to the Sun as the ``elongation angle''), 
even at larger elongation angles it can be comparable in intensity to the isotropic 
(presumably extragalactic) \gray{} background \citep{Abdo2010eg}. 
The flux for
both components of the CR-induced emission is expected to change over
the solar cycle due to the change of the heliospheric flux of the Galactic
CRs in anticorrelation with the variations of the solar activity. 
Observations of the IC emission provide information 
about CR electron spectra throughout the entire inner heliosphere 
simultaneously, 
thus allowing comprehensive studies of the solar modulation of Galactic CRs 
in this region. 
For a moderately high -- moderately low level of the
solar modulation,
the integral IC flux for elongation angles $\le$$6^\circ$
was predicted \citep{Moskalenko2006} to be $F(\ge100\ {\rm MeV}) \sim (2.0-4.3)\times10^{-7}$
cm$^{-2}$ s$^{-1}$, respectively, and thus detectable by the EGRET and \fermilat. 
A calculation by \citet{Orlando2008} gave a similar flux $F(\ge$$100\ {\rm
MeV})=2.18\times10^{-7}$ cm$^{-2}$ s$^{-1}$ for elongation angles $\le$$10^\circ$ for the solar maximum conditions.

Reanalysis of the EGRET data by \citet{Orlando2008} led to the
detection of both predicted components, point-like
hadronic emission from the solar disk and extended leptonic emission from IC scattering of CR electrons on solar photons. 
Their analysis combined all 11 observational periods when the Sun was in the field of view of EGRET 
between the beginning of the \emph{CGRO} mission (April 1991) and the end of the fourth observing cycle (October
1995).
The average solar activity for this period was moderate,
decreasing from its peak in 1990 to the minimum in 1995. The 
\citet{Orlando2008} analysis yielded fluxes
$F(\ge$$100\ {\rm MeV}) = (1.8 \pm 1.1) \times
10^{-7}$ cm$^{-2}$ s$^{-1}$ for 
the disk and $F(\ge$$100\ {\rm MeV})=(3.8 \pm 2.1)\times 10^{-7}$ cm$^{-2}$
s$^{-1}$ 
for the IC component for elongation angles $\le$$10^\circ$,  
consistent with their estimate of the IC flux for the solar maximum conditions.

The launch of \fermi{} in 2008 has made observations of the
quiet Sun with high statistical significance and on a daily basis possible.
During the first two years of the \fermi{} mission, the
solar activity has been extremely low,
resulting in a high heliospheric flux of Galactic CRs. Therefore,
the CR-induced quiescent \gray{} emission from the Sun 
is expected to be near its maximum. 
Preliminary analysis of the 
data from the \fermi{} Large Area Telescope (\fermilat)
showed the existence of both point-like and extended components of solar \gray{} emission \citep{Brigida, Giglietto, Orlando2009}.
In this paper, we report on the \fermilat{} 
observations of the quiescent 
Sun
during the first 18 months of the science phase of the mission.

\section{Data selection and count maps}  \label{sec:data}

\fermi{} was launched on June 11, 2008 into circular Earth orbit with an 
altitude of 565 km and inclination of $25.6^\circ$, and an orbital period 
of 96 minutes. The principal instrument on \fermi{} is the LAT \citep{Atwood}, 
a pair-production telescope with a large effective area
($\sim$8000~cm$^{-2}$ at 1~GeV) and field of view (2.4~sr), 
sensitive to \gray{s} between 20 MeV and $>$300 GeV.
After the commissioning phase, devoted to fine-tuning of the instrument and calibrations,
the \fermilat{} began routine science operations on August 4, 2008. 
The \fermilat{} normally operates in sky-survey mode
where the whole sky is observed every 3 hours (or 2 orbits) with an almost-uniform exposure on daily time scales. 

The energy dependent systematic uncertainties of the effective area of the instrument were evaluated by comparing the 
efficiencies of analysis cuts for data and simulations of observations of
pulsars \citep{Rando2009}. This study revealed a systematic uncertainty of
10\% at 100 MeV, decreasing to 5\% at 560 MeV, and increasing to 20\% at 10 GeV and above. 
The photon angular resolution is also
energy dependent. 
The 68\% containment angle averaged over the \fermilat{} acceptance (the width of the point-spread function -- PSF) 
can be approximated by the following expression:
$\langle\Theta_{68}(\epsilon)\rangle = \left([0.8^\circ \epsilon^{-0.8}]^2 + [0.07^\circ]^2\right)^{1/2}$, 
where $\epsilon$ is the photon energy in GeV.
More details on the instrument performance can be found in the \fermilat{} calibration paper \citep{Atwood}.
The analysis presented here uses post-launch P6V3 instrument response functions (IRFs). 
These take into account pile-up and accidental 
coincidence effects in the detector subsystems that were not considered in the definition of the prelaunch IRFs. 

We use the \fermilat{} data collected between August 4, 2008, and February 4, 2010. 
Events $\ge$100 MeV arriving with elongation angles $\theta\le20^\circ$ 
(region of interest -- ROI) and satisfying the Diffuse class selection
\citep{Atwood} are used. 
To reduce the contamination by the \gray{} emission coming 
from CR interactions in the Earth's upper atmosphere
our selection is refined by selecting
events with zenith angles $<$$105^\circ$.
To reduce systematic uncertainties due to the bright diffuse \gray{} emission from the Galactic plane and a possible spillover 
due to the broad PSF at low energies, 
we have also excluded the data taken when the Sun was within $30^\circ$ of the plane 
($|b_\odot | \ge 30^\circ $). We further excluded the periods when the Sun was within 20$^{\circ}$ of the
Moon or 
any other bright celestial source with the integral flux $F_{\rm 1FGL}\ge2 \times 10^{-7}$ cm$^{-2}$ s$^{-1}$ above 100 MeV as selected 
from the 1FGL \fermilat{} source catalog \citep{Abdo_cat}. These various selections produce a very clean event subsample 
but at the expense of removing about 93\% of the initial ROI data set as summarized in Table~\ref{table4}.  

Because the Sun is moving across the sky, the analysis of its emission requires special treatment. Therefore, 
a dedicated set of tools was developed,
not a part of the standard \fermilat{} Science Tools package,
to deal with moving  
sources such as the Sun and the Moon.
Using these specialized tools, the data are
selected in a moving frame centered on the instantaneous solar position, which is computed using an interface to the 
JPL ephemeris libraries\footnote{http://iau-comm4.jpl.nasa.gov/access2ephs.html}.

\section{Analysis method} \label{sec:analysis}

For the analysis of the Sun-centered maps we used the \fermilat{} Science
Tools\footnote{Available from \fermi{} Science Support Center (FSSC), http://fermi.gsfc.nasa.gov/ssc \label{fssc}} version 9r16p0.
The \emph{gtlike} tool provides maximum likelihood parameter values \citep[using the method described in][]{TS3,TS1,TS2}, 
which derives error estimates (and a full covariance matrix)
from Minuit, a minimization tool supported by CERN, using the quadratic approximation around the best fit.

\subsection{Background determination \label{background}}

The correct evaluation of the background in the region around the Sun is of considerable
importance for the analysis of the weak extended IC emission. 
The latter is expected to decrease as $\sim$$1/\theta$
with elongation angle $\theta$ \citep{Moskalenko2006,Orlando2007} and becomes indistinguishable from the background 
for $\theta\ga20^\circ$.
The background is mainly due to the diffuse Galactic and isotropic (presumably extragalactic) \gray{} 
emission averaged along the ecliptic and to weak point sources. 
The evaluation of the 
background was done by two methods, one based on the analysis of flight data and the other on simulations.

The first method utilizes the data and
is called 
a ``fake-Sun'' analysis, 
where an imaginary source trails the Sun along the ecliptic. 
For this method, application of exactly
the same sets of cuts as are used for the Sun yields an estimate of the background.
Since the extended solar IC emission is insignificant for elongation angles $\theta\ga$$20^\circ$ 
we used $40^\circ$ as the minimum trailing distance.
To reduce statistical errors in the background determination,
the background is averaged over 4 fake-Sun sources displaced from each other and from the Sun itself by 
$40^\circ$ intervals along the ecliptic. 
Because all fake-Sun sources are sampling the same area on the sky, the backgrounds determined
using individual fake-Sun sources are consistent within a fraction of a percent.
Note that the 18-month analysis period is long enough to average out any effects connected with incomplete
sampling of the background.
This would be an issue if the analysis period was shorter than 12 months.

\begin{figure}[!t]
\centering
\includegraphics[width=0.5\textwidth]{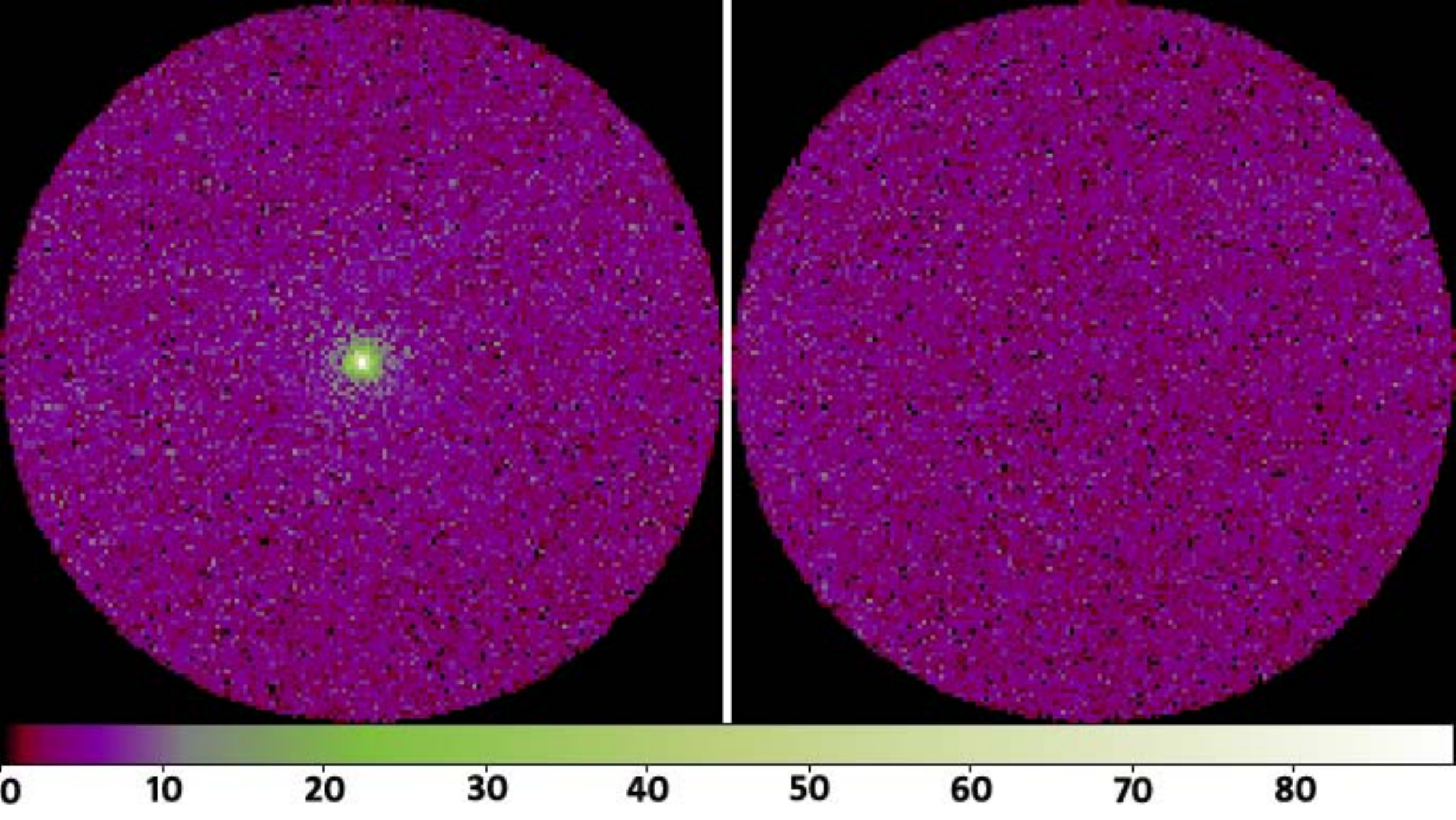}
\caption{Count maps for events $\ge$100 MeV taken between August 2008 and February 2010 and centered on the Sun (left) and on the trailing source (so-called fake-Sun, right) representing the background. The ROI has $\theta=20^\circ$ radius and pixel size $0.25^\circ \times 0.25^\circ$. The color bar shows the number of counts per pixel. 
\label{fig:map}}
\end{figure}

\begin{figure}[!t]
\centering
\includegraphics[width=0.5\textwidth]{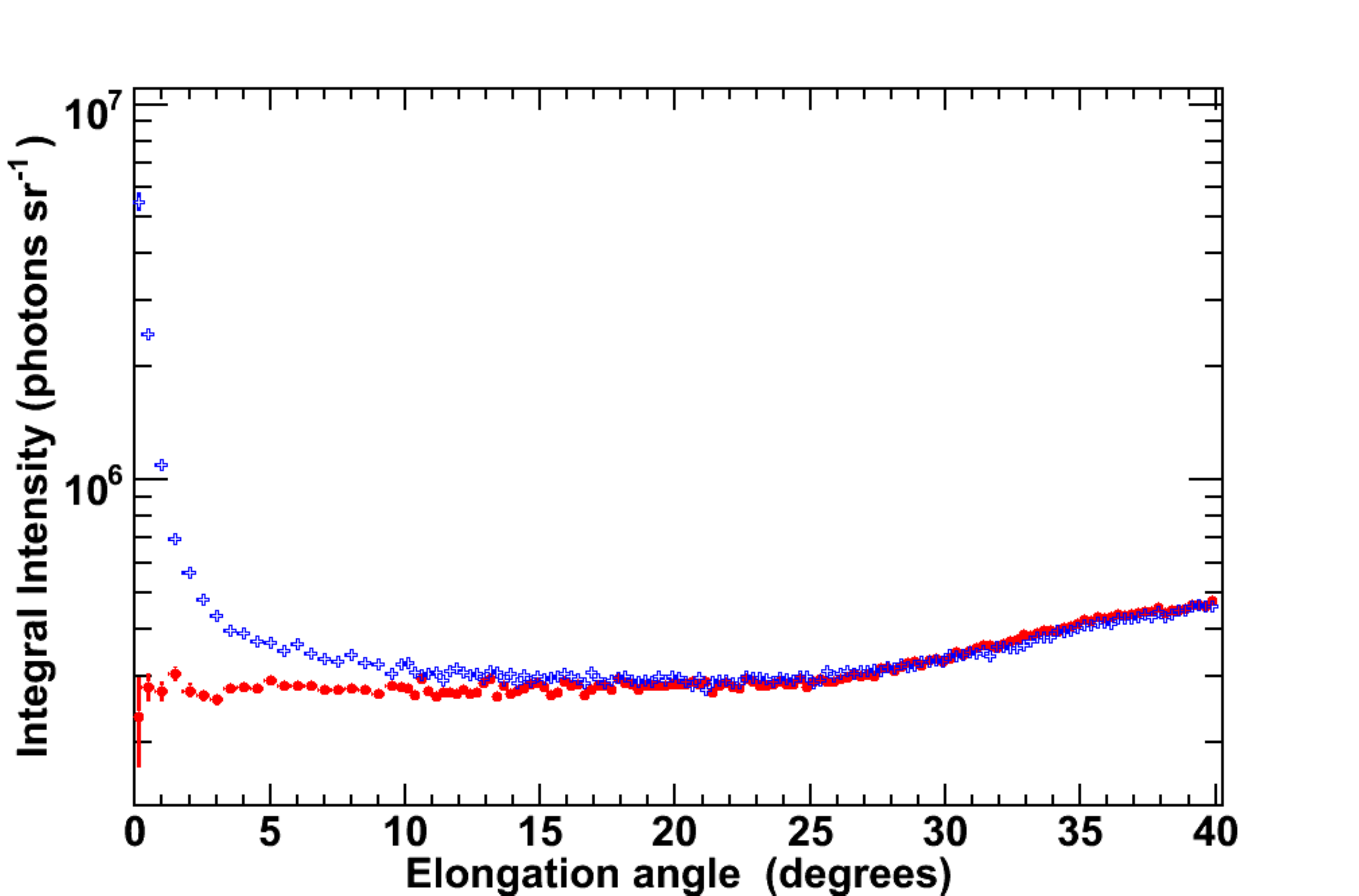}
\caption{ 
Integral intensity ($\ge$100 MeV) plot for the Sun-centered sample vs.\ elongation angle, bin size: $0.25^\circ$.
The upper set of data (open symbols, blue) represents the Sun, the lower set of data (filled symbols, red) represents 
the ``fake-Sun'' background.}
\label{fig:fig3}
\end{figure}

Figure~\ref{fig:map} shows the Gaussian-smoothed count maps
$>$100 MeV centered on the solar position and the hypothetical trailing source 
(average of the 4 fake-Suns).
The solar emission is clearly seen on the left panel, while the right panel shows the background, which is essentially uniform. 
The 
integral intensity distribution for the two samples, centered at the solar position and centered on the averaged fake-Sun source,
is shown in Figure~\ref{fig:fig3}. The number of events per solid angle is shown vs.\ the 
angular distance from the Sun (the elongation angle) and the fake-Sun positions for a bin size $0.25^\circ$. 
While for the solar centered data set the 
integral intensity
increases considerably for small
elongation angles, the averaged fake-Sun profile is flat.  
The two distributions overlap at distances larger than $20^\circ$ where the signal significance is diminished.
The gradual increase in the integral intensity for $\theta\ga$$25^\circ$ is due to the bright Galactic plane broadened by the PSF,
see the event selection cuts summarized in Section~\ref{sec:data} and Table~\ref{table4}.

\begin{figure}[!t]
\centering
\includegraphics[width=0.5\textwidth]{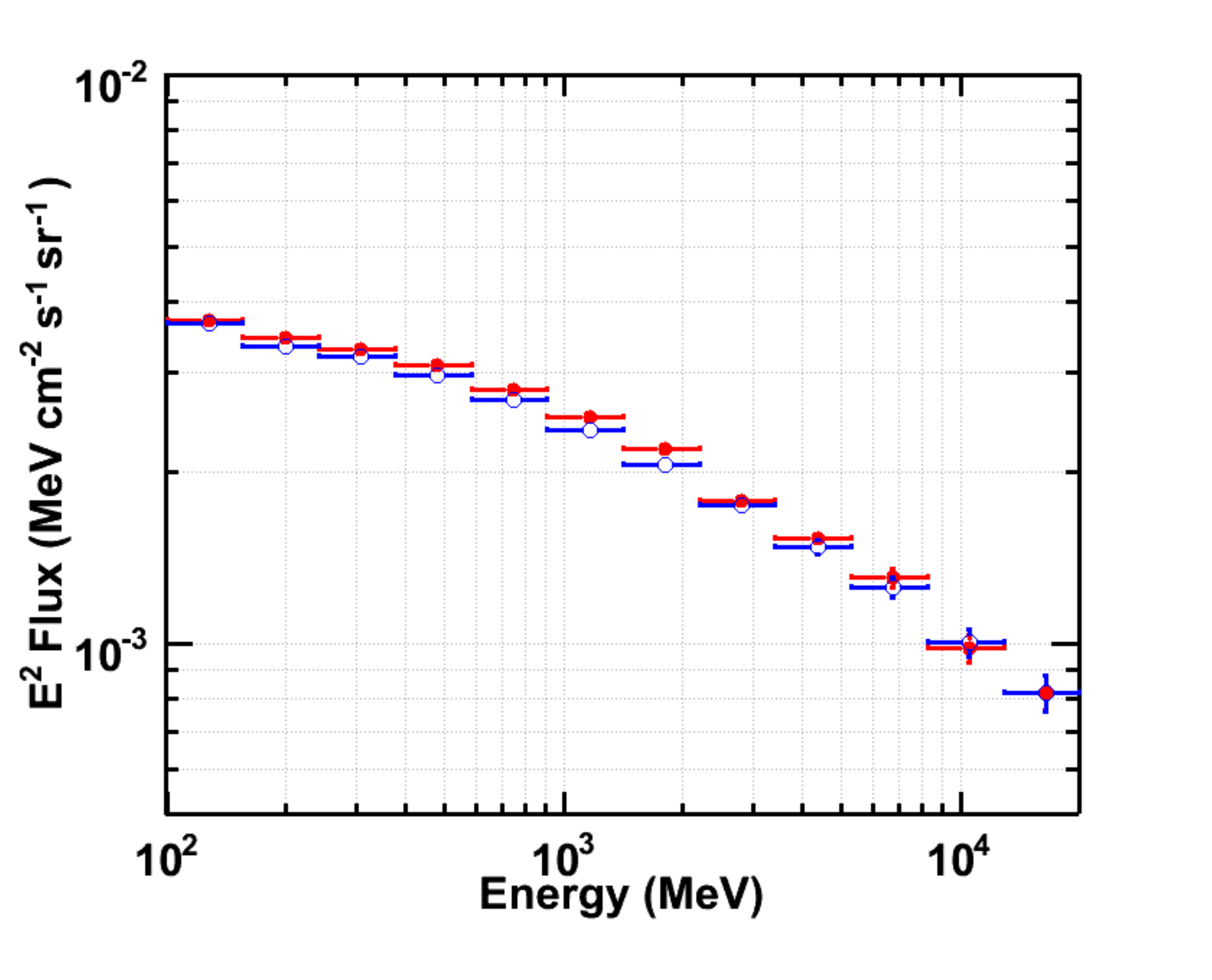}
\caption{
Reconstructed spectrum of the background for the fake-Sun method (filled symbols, red) and for the simulated background sample (open symbols, blue) averaged over a 20$^\circ$ radius around the position of the Sun.
}
\label{fig:fig4}
\end{figure}

The second method of evaluating the background uses an all-sky
simulation which takes into account a model of the
diffuse emission (including the Galactic and isotropic components, 
gll\_iem\_v02.fits and isotropic\_iem\_v02.txt, correspondingly; see footnote \ref{fssc}) 
and the sources from 1FGL \fermilat{} catalog \citep{Abdo_cat}. 
To the simulated sample we apply the same set of cuts as applied to the real data and
select a subsample centered on the position of the real Sun.
The simulated background is then compared with the background derived from 
a fit to the fake-Sun in the first method.  
Figure~\ref{fig:fig4} shows the spectra of the
background derived by the two methods. 
The agreement between the two methods \citep[and the spectrum of the diffuse emission at medium and high latitudes,][not shown]{Abdo2010eg} is very good, 
showing that the background estimation is well understood and that there is no
unaccounted or missing emission component in the analysis.

Finally, we check the spatial uniformity of the background determined by 
the fake-Sun method.
The ROI restricted by $\theta\le 20^\circ$ was divided into nested rings.
We use 4 annular rings with radii $\theta=10^\circ$, $14^\circ$, $17.3^\circ$,
and $20^\circ$, which were chosen to subtend approximately the
same solid angle for each ring, and hence should 
contain approximately equal numbers of background photons if their 
distribution is spatially flat. The ring-by-ring background 
intensity variations were found to be less than 1\%. 
Note that 
the background emission is considerably more intense than the expected IC component (see Section~\ref{separation}), and
even small background variations across the ROI may affect the analysis 
results. To minimize these systematic errors, we therefore using the ring method for the background evaluation.

\begin{figure}[t!]
\centering
\includegraphics[width=0.5\textwidth]{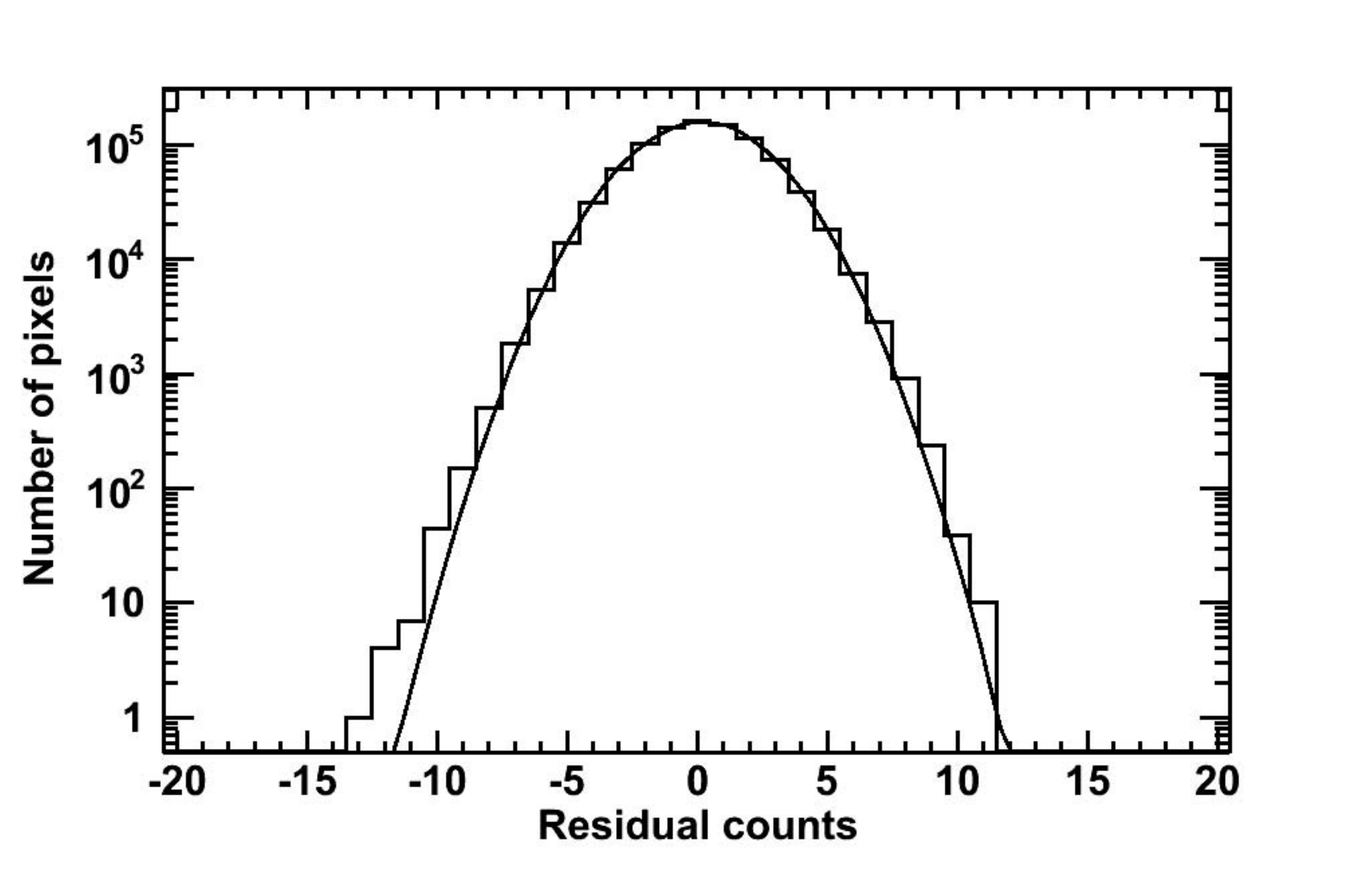}
\caption{Distribution of the residual counts: differences between the photon counts in the map centered at the average fake-Sun and the 
high statistics count map derived from the model. A normal distribution
with the same parameters is shown by the bold line.
The map used to generate this distribution
has a bin size of $0.3^\circ$ centered on the solar position of the 
simulated sample.
}
\label{fig:fig5}
\end{figure}

The evaluated spectrum of the background for 
$\theta\le20^\circ$ was fitted using the maximum
likelihood method and the results were used to derive the simulated average photon 
count per pixel using the \emph{gtmodel} routine from the \fermilat{} Science Tools.
For each pixel of the predicted average photon count distribution, we generate
a set of 100 random events assuming Poisson statistics and 
compare them with the
observed number of photons in the corresponding pixel of the map centered on
the fake-Sun, as shown in Figure~\ref{fig:fig5}.
The resulting distribution of the difference of counts between the observed and simulated photons per pixel is consistent with a
normal distribution of mean $0.155 \pm 0.002$ and standard deviation $2.337 \pm 0.002$.

\subsection{Separation of the solar emission components} \label{separation}

The separation of the disk and extended components of the 
solar emission was done in a model-independent way. 
The Sun-centered maps of \fermilat{} \gray{} counts were analyzed using a maximum likelihood technique where the background
parameters in each nested ring were fixed using the fake-Sun method as described above. 
The flux and spectral index of the extended emission were determined independently in each energy range
in each nested ring, while the disk component (modeled as the \fermilat{} PSF 
because the $\sim 0.5^\circ$ solar disk is not resolvable) 
was allowed to have a free spectral index and flux normalization.
The choice of the annular radii ($5^\circ$, $11^\circ$, and $20^\circ$) has been optimized to have a likelihood test 
statistic\footnote{The likelihood test statistic TS is defined in \citet{TS2}.}
(TS) value $\ge25$ for the fitted IC component in each ring.

\begin{figure}[!t]
\centering
\includegraphics[width=0.5\textwidth]{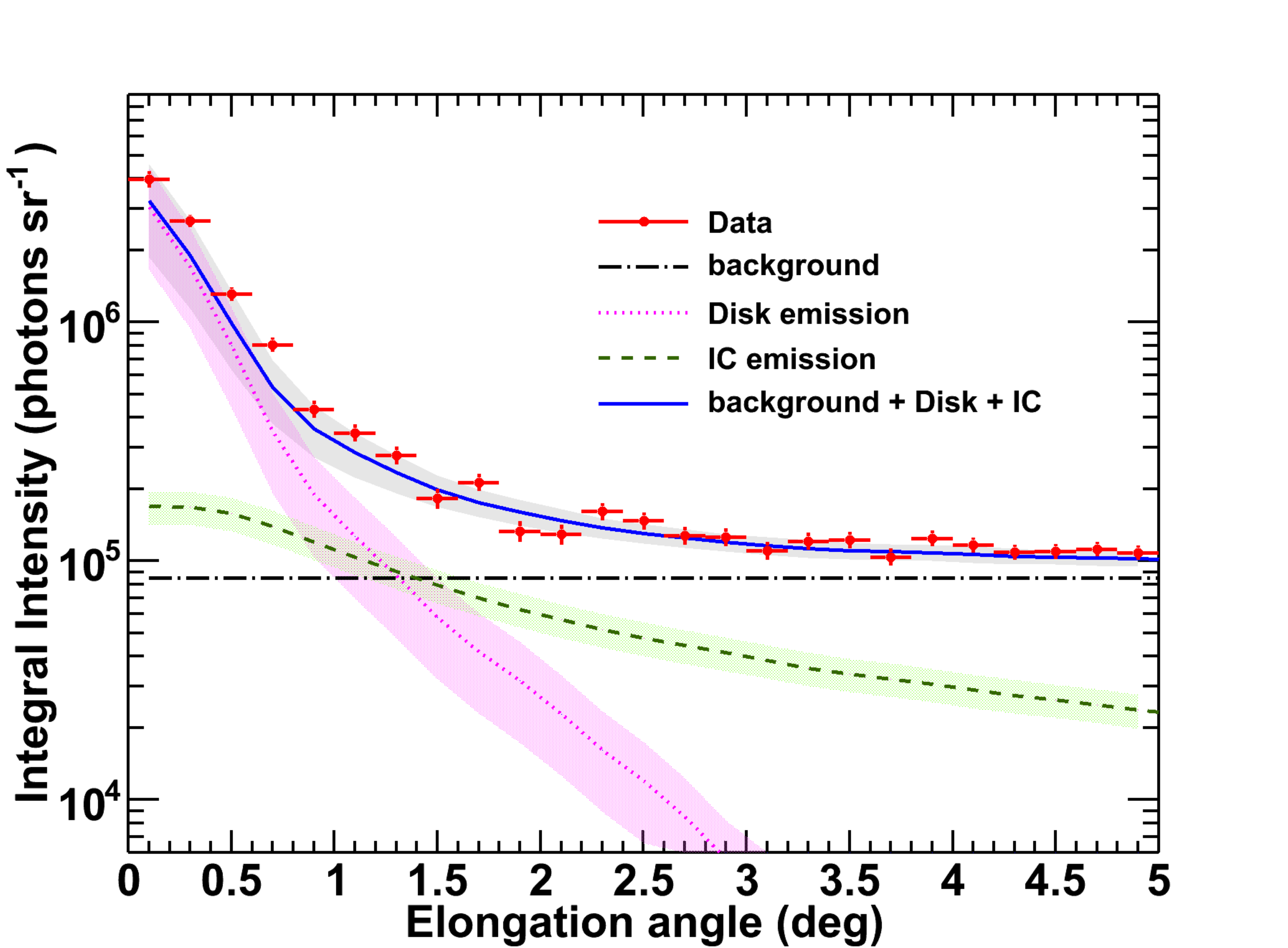}\medskip\\
\includegraphics[width=0.5\textwidth]{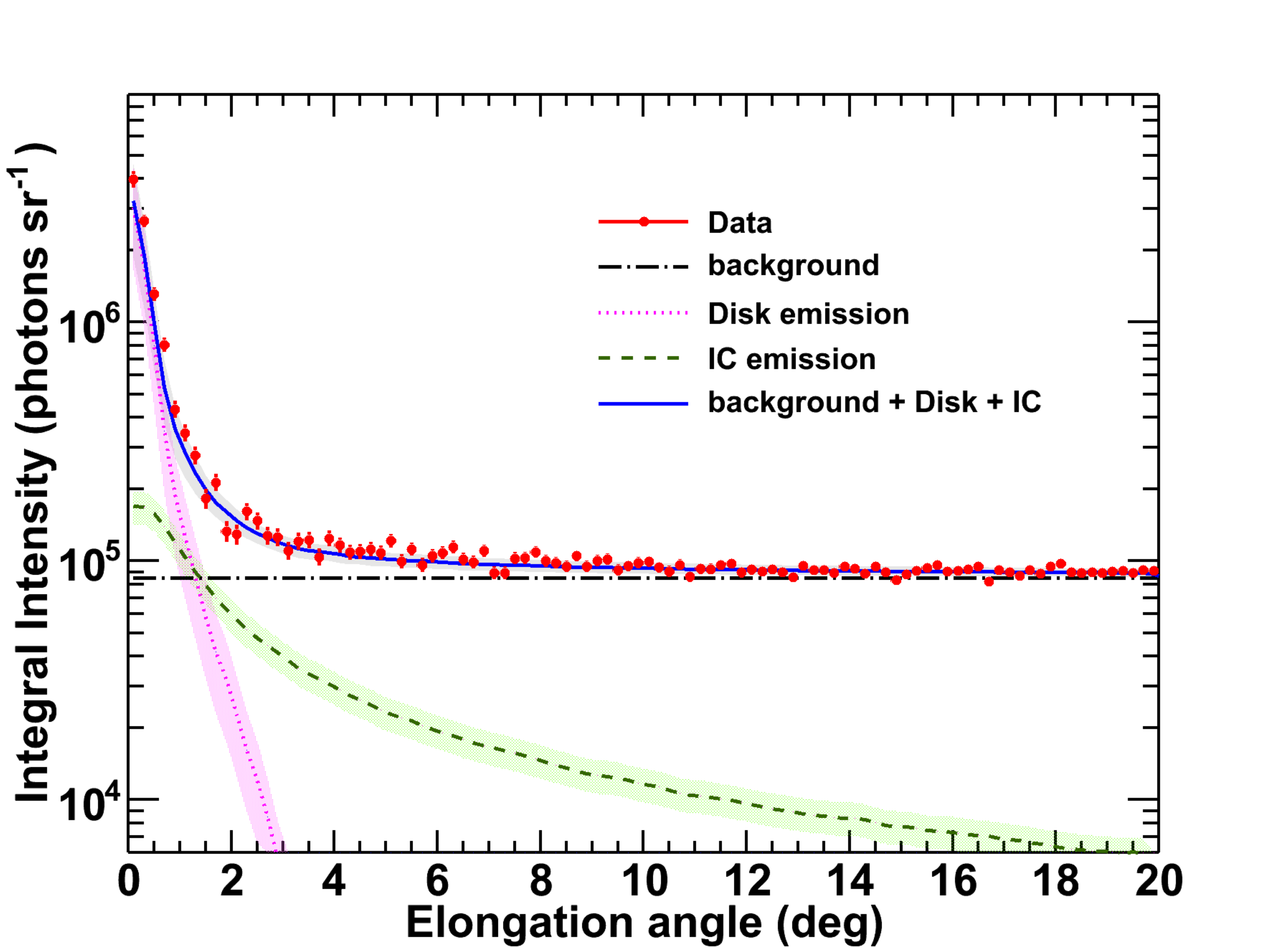}
\caption{ 
{Integral intensity} profiles above 500 MeV for elongation angles $\le$$5^\circ$ (top) and $\le$$20^\circ$ (bottom). 
Points (red) are the observed counts, dash-dotted horizontal (black) line is the background, dotted (magenta) and dashed (green) lines are the point-like and extended components of the emission, correspondingly. The solid (blue) line is the sum of the background and
the two components of the emission. The shaded areas around the lines show total error estimates. See text for details. 
}
\label{fig:fig6}
\end{figure}

Figure~\ref{fig:fig6} shows the angular distribution of photons $\ge$500~MeV 
from the Sun on a $0.2^\circ$ grid,
the background determined by the fake-Sun method, and the fitted 
disk and extended components. 
The spectrum of a disk 
source is modeled as a power law with the total 
flux and spectral index obtained from the fit of the disk component. 
The extended emission is modeled as the sum of the individual fits in each energy bin in each nested ring.
The observed angular distribution can be well fitted only by adding an extended IC component.

\section{Results}\label{sec:results}

The analysis of the \fermilat{} observations for the first 18 months of 
the mission gives a significant 
detection and separation of the two components of
solar \gray{} emission.
The large photon statistics allow us to derive the spectral shape of each component 
by fitting them in narrow energy bands, 
so that the exposures and convolutions with the 
PSF do not depend strongly on a the assumed spectral shape.

Table~\ref{table2} gives 
the energy spectrum for the IC component from
the model-independent analysis 
or elongation angles $\le$$5^\circ$, 
which corresponds to the radius of the innermost ring, and $\le$$20^\circ$. 
The fitted integral fluxes for the IC component in each ring are given in Table~\ref{table1}
together with model calculations (described in Section~\ref{sec:ic}).
The energy bin size was selected to provide 
good convergence of the likelihood fit in each energy interval.
The observed spectrum for the disk component is given in Table~\ref{table3}. 

Evaluation of the systematic errors for each component of the solar emission has to take into 
account the uncertainties 
in the effective area as a function of energy as indicated in Section~\ref{sec:data} and
the statistical uncertainties in the determination of the background (Section~\ref{background}). 
The uncertainties associated with the effective area were propagated using the modified effective areas 
bracketing the nominal ones (P6V3 Diffuse) \citep{Abdo2010crab}. 
The uncertainties in the background were taken into account
by assigning $\pm$$1\sigma$ deviation to the background flux and repeating the fit for all components.

\section{Calculation of the IC emission} \label{sec:ic}

The calculation of the IC emission from the solar radiation field was 
first described by \citet{Moskalenko2006} and
\citet{Orlando2008} using the formula for the
differential interaction rate for an
anisotropic
distribution of target photons
\citep{Moskalenko2000}. 
The CR electron distribution was assumed to be isotropic 
(hereafter we use the term electrons to refer to both
electrons and positrons).
The upscattering of optical
solar photons to the \fermilat{} energy range $\ge$100 MeV
involves CR electrons above $\sim$2 GeV. At energies
below $\sim$20 GeV, CR electrons  in the heliosphere are subject to
significant adiabatic energy losses
and drifts in the magnetic field: the
combined effect is called the heliospheric modulation \citep[e.g.,][]{Potgieter1998}.
A calculation of the spatially extended emission due to the IC scattering of CR electrons on solar photons requires 
the integration of \gray{} yields along the line-of-sight folded with the CR electron spectrum at different 
heliospheric distances \citep{Moskalenko2006}.

Due to the radial distribution of the solar photons the IC emission is brightest at small elongation angles,
and is distributed over the whole sky at low levels.
The differential flux of upscattered photons depends on how the CR electron
spectrum
changes with heliocentric distance, which is the only unknown in the calculations. Therefore, observations of the IC component
of the solar emission provide a new tool to probe the CR electron spectrum 
in the inner heliosphere ($r<10$ AU) down to close proximity
to the Sun.

\subsection{CR modulation in the heliosphere}

Studies of the CR modulation are based on the solution of the \citet{Parker1965} transport equation \citep[e.g., 
see reviews by][]{Potgieter1998,Heber2006}. Particle transport to the inner
heliosphere is mainly determined by spatial diffusion,
convection by the solar wind, and adiabatic cooling. Besides, rotation of the Sun causes
the interplanetary magnetic field in the solar equatorial plane to be distinctly different from the field above and below the poles,
the so-called Parker's spiral \citep{Parker1958}. 
This, in turn, causes differences in the spectra of CR particles in these regions.
Realistic time-dependent three-dimensional hydrodynamic models incorporating
these effects have been developed \citep[e.g.,][]{Florinski2003,Langner2006,
Potgieter2004}; however, the effect of heliospheric modulation is still
far from being fully understood. 
One of the major difficulties in developing models of heliospheric modulation is that the data
gathered by spacecraft 
taken at different heliospheric distances are often at different energies and related to different modulation levels.
The region $<$1 AU is the least studied. 
Another problem is that the input information for the modulation models, such 
as the local interstellar spectra of CR species, is missing.

Although much effort has gone into development of realistic modulation models,
the method most often 
used is the so-called ``force-field'' approximation \citep{gleeson}, which employs
a single parameter -- the ``modulation potential'' $\Phi$ -- that varies over the solar 
cycle to characterize the strength of the modulation effect on the CR spectra:

\noindent
\begin{equation}
J_e(r,E)=J_e\left( \infty, E+ \Phi[r] \right)  \frac{(E^2- m_e^2 c^4)} {(\{E+\Phi[r]\}^2-m_e^2 c^4)},
\label{forcefield}
\end{equation}

\noindent
where $J_e ( \infty, E+ \Phi[r])$ is the local
interstellar electron spectrum, $E$ is the total
electron energy, $m_ec^2$ is the electron rest mass,
$\Phi(r)$  is the modulation potential, and $r$ is the
distance from the Sun. Even though this approximation is very crude 
and implies spherical symmetry for the heliosphere, it can be used 
as a convenient \emph{parameterization} of the CR spectrum at different levels of solar activity.

Expressions for $\Phi(r)$ were derived by \citet{Moskalenko2006} based
on the radial 
dependence of the CR mean free path given by \citet{fujii} for Cycles 20/22 and and for Cycle 21 assuming separability of the heliospheric diffusion coefficient into radial and energy-dependent functions. 
For Cycles 20/22, \citet{Moskalenko2006} obtain:

\noindent
\begin{equation}
\Phi_1(r)= \frac{\Phi_0}{1.88}\left\{
\begin{array}{ll}
r^{-0.4} - r_b^{-0.4}, & r\ge r_0,\\
0.24 + 8 (r^{-0.1} - r_0^{-0.1}), & r<r_0, 
\end{array}
\right.
\label{eq7}
\end{equation}

\noindent
where $\Phi_0$ is the modulation potential at 1 AU, $r_0=10$ AU, and $r_b=100$ AU is the heliospheric boundary. For Cycle 21 they give:

\noindent 
\begin{equation}
\Phi_2(r)= \Phi_0 (r^{-0.1} - r_b^{-0.1}) / (1 - r_b^{-0.1}).
\label{eq8}
\end{equation}

\noindent
These formulae were derived for $r\ga1$ AU. Closer to the Sun at 0.3 AU $\la r<1$ AU there
are only very few measurements of 
Galactic CR protons and helium, and these are at low energies $\la$60 MeV nucleon$^{-1}$ \citep[e.g.,][]{Christon1975,Kunov1977,Muller1977}.
Since CR transport is strongly energy dependent, these low-energy protons and helium nuclei measurements are 
irrelevant to the current study and CR transport is very uncertain. 
Therefore, as a first approximation we use eqs.~(\ref{eq7}), (\ref{eq8}) for the entire heliosphere (Models 1, 2),
from the solar surface to the heliospheric boundary $R_\sun < r < r_b$, except for Model 3 described below.

\subsection{The CR electron spectrum}

Assuming that the CR propagation in the heliosphere is spherically symmetric in the first approximation,
the CR electron spectrum at different heliospheric distances is the only unknown in evaluating the solar IC flux. 
Comparison of the model calculations with 
observations provides a method to probe the CR electron spectra at
different heliospheric distances $r\la10$ AU. 
Because of the lack of CR measurements outside of the heliosphere, we have to rely on a comprehensive 
Galactic CR propagation model tuned to other CR data and diffuse \gray{} emission, such as GALPROP \citep{SM1998,Moskalenko1998, Ptuskin2006,Abdo2009diffuse,Abdo2010eg,AV2010}, 
or measurements of the flux at 1 AU. 

In this paper we use the CR electron spectrum
recently measured by the \fermilat{} \citep{fermi_el, Ackermann2010} between 7 GeV and $\sim$1 TeV.
At energies $<$7 GeV we use the Alpha Magnetic Spectrometer (AMS-01) lepton data \citep{AMS2000leptons}
collected in its flight in June 1998, which are the most accurate to date, but 
were made in
the previous solar cycle. Preliminary results by the Payload for Antimatter Matter Exploration and Light-nuclei Astrophysics
\citep[PAMELA,][]{PAMELA2011electrons}, in orbit since June 2006,
indicate that the electron spectrum $>$7 GeV is consistent with the \fermilat{} data, but is
more intense than the AMS-01 data at lower energies.

Unusually high intensities of Galactic CRs during the period of anomalously low solar activity
were also reported by the 
Advanced Composition Explorer (ACE), which monitors the flux of CR species at 1 AU in a few 100 MeV/nucleon energy range.
The measured intensities of major species from C to Fe \citep{Mewaldt2010} were each 20\%-26\% greater in late 2009 than
in the 1997-1998 minimum and previous solar minima of the space age (1957-1997). 
While the value of $\Phi_0$ is somewhat arbitrary and depends on the assumed local interstellar spectrum (but does not depend on the particle type), the unusually
low solar activity during the observational period suggests that the modulation of Galactic CRs should be considerably weaker than at any other time.

Therefore, the \fermilat{} CR electron spectrum was fitted using a parameterization
for the \emph{local interstellar spectrum} and assuming a relatively low
value of the modulation potential $\Phi_0=\Phi(1\ {\rm AU})=400$ MV (eq.~[\ref{forcefield}]):

\noindent
\begin{equation}
J_e(\infty
,E)= \left\{
\begin{array}{ll}
& a (b+c)^{-d} (E/b)^{-3}, E<b,\nonumber \\
& a(E+c)^{-d}, E\ge b, \nonumber
\end{array}
\right.
\label{fermi_e1}
\end{equation}

\noindent
where
$J_e$ is in units m$^{-2}$ s$^{-1}$ sr$^{-1}$ GeV$^{-1}$,
$a = 160.24$ m$^{-2}$ s$^{-1}$ sr$^{-1}$ GeV$^{2.03}$,
$b = 7$ GeV,
$c = -1.20$ GeV, 
$d = 3.03$,
and $E$ is the total electron energy in GeV. 
Different values for $\Phi_0$ yield somewhat different values for the other
parameters while providing the same fit to the \fermilat{} electron spectrum. 
We found that the value of $\Phi_0$ given above provides reasonable agreement with the observed IC flux 
(see Section~\ref{sec:results}); more detailed analysis of the electron spectrum in the 
inner heliosphere based on \fermilat{} observations of the IC emission will be given in a forthcoming paper.

Since most of the IC emission is produced by
electrons in the inner heliosphere 
$r<10$ AU, eqs.~(\ref{forcefield}), (\ref{fermi_e1}) 
is a good approximation of the CR electron spectrum even though it may produce an unphysical 
spectrum beyond the heliospheric boundary at $r_b=100$ AU.
(In fact, for elongation angles $\theta\le20^\circ$ most of the emission is produced by electrons at $r\la2$ AU, see Figure 2
in \citealt{Moskalenko2006}.)
However, the electron spectrum obtained in such a way is also close to the local
interstellar spectrum that 
is calculated by GALPROP and used for modeling of the Galactic diffuse \gray{}
emission \citep{Abdo2009diffuse,Abdo2010eg}.

Using this parameterization for the CR electron spectrum, we construct three
models to compare with the data. 
To obtain the electron spectrum at an arbitrary heliospheric distance, $r$,
Models 1 and 2 use parameterizations $\Phi_{1,2}(r)$ (with $\Phi_0=400$ MV)
given by eqs.~(\ref{eq7}), (\ref{eq8}), respectively. 
Model 3 uses parameterization eq.~(\ref{eq7}) as specified above, but assuming \emph{no additional} modulation for $r<1$ AU, i.e., $\Phi(r<1\ {\rm AU})=\Phi_0$. In this model the electron spectrum at $<$1 AU is the same as measured by the \fermilat, which provides us with an upper limit for CR electron flux closer to the Sun. 

\begin{figure}[!t]
\centering
\includegraphics[width=0.5\textwidth]{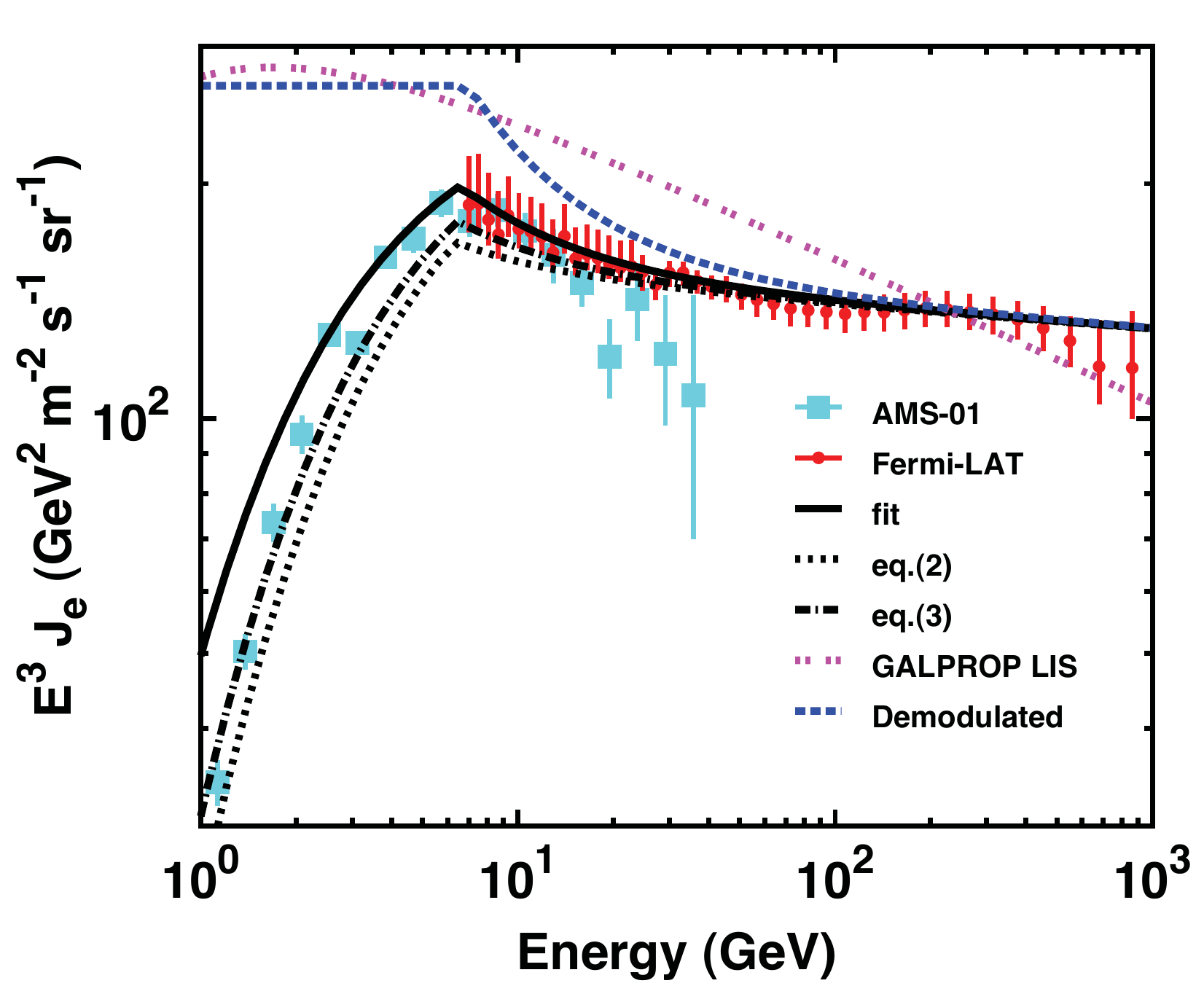}
\caption{Electron spectrum used in the models of the IC emission compared with the data from \fermilat{} \citep[small red filled circles,][]{fermi_el, Ackermann2010} and AMS-01 \citep[blue filled squares,][]{AMS2000leptons}. The thick solid (black) line is a fit to the \fermilat{} electron spectrum at 1 AU; this is also the electron spectrum for $r<1$ AU in Model 3. The dash-dotted (black) and doted (black) lines show the electron spectra at at $r=0.3$ AU in Models 1 and 2 calculated for $\Phi_0=400$ MV, eq.~(\ref{eq7}) and eq.~(\ref{eq8}), correspondingly. The dashed (blue) line shows the demodulated local interstellar spectrum. The double-dot (magenta) line shows the local interstellar spectrum of leptons as calculated by GALPROP \citep{Ptuskin2006}; this spectrum is used in the model calculations of the diffuse Galactic emission.    
\label{CRE}}
\end{figure}

Figure~\ref{CRE} shows the fit to the \fermilat{} electron spectrum. 
The AMS-01 lepton data were collected during the period of rising solar 
activity\footnote{See the charts of the computed tilt angle of the heliospheric current sheet at http://wso.stanford.edu/}, where the 
low modulation potential adapted for the current solar minimum is not appropriate. 
These data are not fitted and shown for only illustration.
To illustrate the effect of different parameterizations $\Phi_{1,2}(r)$,
the Figure also shows the spectra for Models 1, 2
at $r=0.3$ AU. 
All three models of the electron spectrum are normalized to the same \fermilat{} data at 1 AU. Therefore the largest difference
between the models is at low energies where the modulation effect is strongest. 

\section{Emission from the solar disk} \label{sec:disk}
 
A detailed theoretical study of \gray{} emission from the solar disk
was published by \citet{Seckel1991}.
The predicted flux is very sensitive to the assumptions about CR transport in the inner heliosphere, the magnetic fields in the solar atmosphere, and the CR cascade development in the solar atmosphere.  

The \citet{Seckel1991} calculation divides space into three regions:
the interplanetary space, the corona, and the regions below the corona.
The interplanetary magnetic field ($>$$2R_\odot$) is taken to be in the form of
a \citet{Parker1958} spiral with the total field $\sim$50 $\mu$G near the Earth.
Inferior to Earth's orbit the field is nearly radial at angle $\sim$$45^\circ$.
CR propagation from the Earth's orbit to the bottom of the corona is treated as
spherically symmetrical diffusion with no absorption (i.e., neglecting CR interactions). The adopted
diffusion coefficient has a linear dependence on energy, while the spatial
radial dependence is a power-law with index 2. The CR spectrum at 1 AU was
adopted from \citet{Webber1989}, which is a parameterization of the measurements made
during balloon flights in 1976 and 1979 \citep{Webber1987}.

Interior to the corona, the CR propagation and absorption (interactions) are treated simultaneously using a Monte Carlo code.
The magnetic field configuration is chosen corresponding to a quiet Sun \citep{Priest1982}, where the average field strengths are of the order of a few Gauss. However, the fields are non-uniform with the flux bundles ($\sim$$10^3$ G, a few hundred kilometers across) located at the corners of convective cells and separated by thousands of kilometers.  
The chromosphere is assumed to be isothermal in hydrostatic equilibrium which gives an exponential 
density profile with scale height $\sim$115 km. The density profile below the photosphere is taken from \citet{Baker1966}. The characteristic column density is $\sim$40 g cm$^{-3}$ at 500 km depth and 100 g cm$^{-3}$ at 900 km. 

The cascades initiated by high-energy particles ($>$$3$ TeV) do not contribute much to the observed 
\gray{} flux and are neglected.
The solar magnetic fields do not significantly affect their 
directionality until particle energies drop below $\sim$10 GeV -- by that time most
cascades are deep enough so that only a few low-energy photons will escape. 
On the other hand, the low-energy primaries produce cascades for which a significant
amount of energy is reflected back to the solar surface (so-called mirrored showers).
The typical low-energy cascade has less than a few interaction lengths of
material to pass through before photons can escape. Therefore, such
cascades act as if they evolve in moderately thin targets.

The actual yields are calculated by propagation of one-dimensional cascades
through a slab. The photon yield includes only the photons that make it through
the slab. Secondary electrons, positrons, and 
baryons exiting the slab are ignored even
though they are likely to
reenter the Sun. This may underestimate the actual \gray{} flux. 

The \citet{Seckel1991} calculations were made in two scenarios: (i) neglecting the effects of interplanetary magnetic field on particle propagation and 
assuming the solar surface is fully absorbing (so-called ``naive'' model), and (ii) the ``nominal'' model, which includes all the assumptions
about CR diffusion in the inner heliosphere and corona.  
The integral flux above 100 MeV was predicted to be $F( \ge100\ {\rm MeV}) \sim (0.22-0.65)\times10^{-7}$ cm$^{-2}$ s$^{-1}$ for the ``nominal'' model, where the range corresponds to the different assumptions about the CR cascade development:
slant vs.\ more realistic mirrored showers (i.e., reflected back to the solar surface).

\section{Discussion and conclusions}\label{discussion}

\begin{figure}[!t]
\centering
\includegraphics[width=0.5\textwidth]{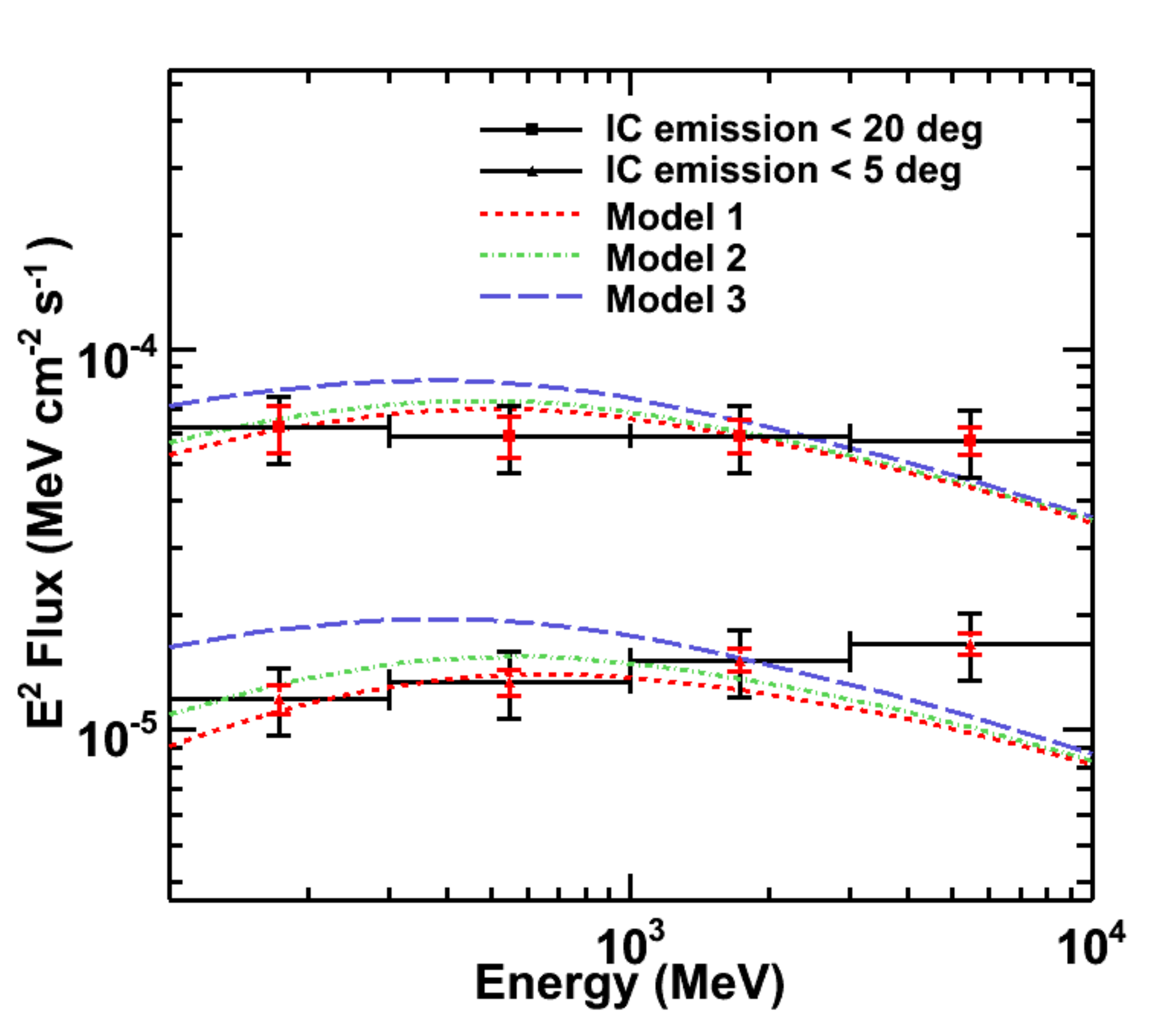}
\caption{Energy spectra of the IC emission for
elongation angles $\le$$5^\circ$ and $\le$$20^\circ$ as observed by \fermilat{} and compared with model predictions. 
Statistical error bars (larger) are shown in black; systematic errors (smaller) are red.}
\label{fig:fig7}
\end{figure}

\begin{figure}[!t]
\centering
\includegraphics[width=0.5\textwidth]{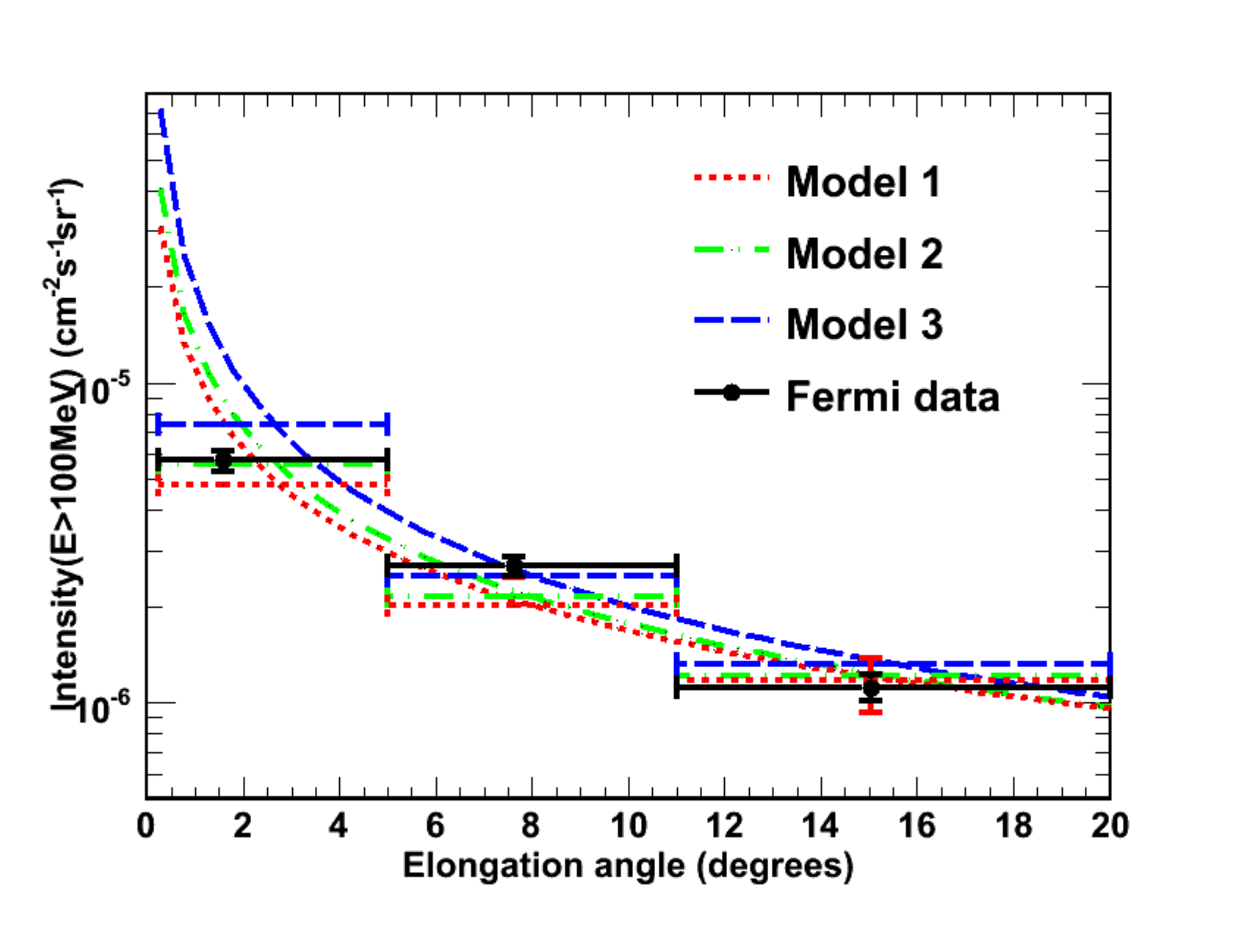}
\caption{Intensity profile for the IC component vs.\  elongation angle compared with the model predictions.
Statistical error bars (smaller) are shown in black; systematic errors (larger) are shown in red. 
To allow a direct comparison with the models, the model predictions are also shown binned with the same 
bin size as used for data. 
}
\label{fig:fig8}
\end{figure}

Figures~\ref{fig:fig7} and \ref{fig:fig8} show the results of the analysis of the extended emission component together
with the model predictions (Section~\ref{sec:ic}).
The plotted values are obtained from the IC flux in each ring as shown in Table~\ref{table1} 
divided by the corresponding solid angle. The model calculations
are shown unbinned (curves) and binned with the same bin size as used for the data.
Although the highest energy point $3-10$ GeV shows some excess relative to the model predictions, 
this is difficult to explain from the model viewpoint since the effect of the solar modulation is decreasing at high energies thus
making the model more accurate. Future analysis with larger statistics should clarify if this discrepancy is real. 
The agreement of the observed spectrum and the angular profile of 
the IC emission with the model predictions (as described in Section~\ref{sec:ic}) 
below a few GeV is very good.
The innermost ring used for the analysis of the IC emission 
subtends an angular radius  of $5^\circ$ 
corresponding to a distance $\sim$0.1 AU from the Sun, i.e.\ 4 times closer to the Sun than Mercury. 
At such a close proximity to the Sun, and actually anywhere $<$1 AU, the spectrum of CR electrons has never been measured. 

It does not seem possible to discriminate between the models at the current stage. The spectral shape $<$1 GeV in 
Figure~\ref{fig:fig7} and the intensity in the innermost ring in Figure~\ref{fig:fig8} is better reproduced by Models 1, 2, while the intensity
in the middle ring $5^\circ-11^\circ$ (Figure~\ref{fig:fig8}) is better reproduced by Model 3.
Even though the current data do not allow us to discriminate between different models of the CR electron 
spectrum at close proximity to the Sun, the described analysis demonstrates how the method would work once the 
data become more accurate. In particular, it is possible to increase the statistics by four-fold by  
masking out
the background sources or modeling them, instead of requiring the angular separation between bright sources and the Sun to be $>$$20^\circ$ (Table~\ref{table4}). More details will be given in a forthcoming paper. 
The increase of the solar activity may also present a better opportunity to distinguish between the models 
since the difference between the model spectra of CR electrons will increase with
solar modulation.

The intensity of the IC component is comparable to the intensity of the isotropic \gray{} background even for 
relatively large elongation angles (Table~\ref{table2}).
Integrated for subtended angles $\le$$5^\circ$, the latter yields $\sim$$2.5\times10^{-7}$ 
cm$^{-2}$ s$^{-1}$ above 100 MeV \citep{Abdo2010eg} vs.\ $\sim$$1.4\times10^{-7}$ 
cm$^{-2}$ s$^{-1}$ for the IC component. For subtended angles $\le$$20^\circ$, the 
integral flux of the isotropic \gray{} background is 
$\sim$$3.9\times10^{-6}$ cm$^{-2}$ s$^{-1}$ above 100 MeV vs.\ $\sim$$6.8\times10^{-7}$ cm$^{-2}$ s$^{-1}$
for the IC component. 
Therefore, it is important to take into account the broad nonuniform IC component of the solar emission when 
dealing with weak sources near the ecliptic. The relative importance 
of the IC component will increase with time since the upper limit on the 
truly diffuse extragalactic emission could be lowered in future as more \gray{} sources are discovered
and removed from the analysis.

\begin{figure}[!t]
\centering
\includegraphics[width=0.5\textwidth]{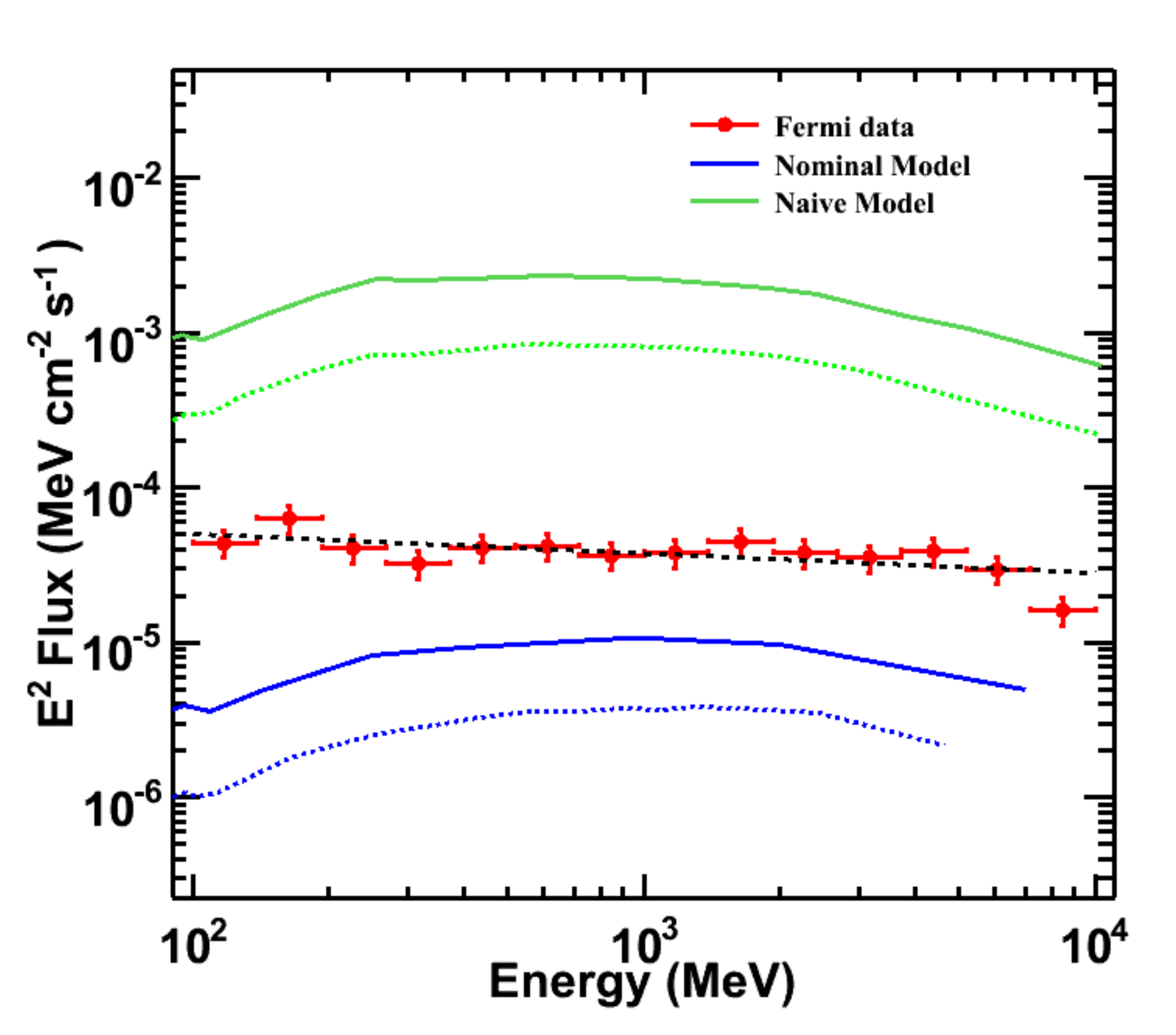}
\caption{Energy spectrum for the disk emission as observed by the \fermilat. The curves show 
the range for the ``nominal''  (lower set, blue) and ``naive''  (upper set, green) model predictions by 
\citet{Seckel1991} for different assumptions about CR cascade
development in the solar atmosphere (see text for details). 
The black dashed line is the power-law fit to the data with index $2.11\pm0.73$. 
}
\label{fig:fig9}
\end{figure}

Figure~\ref{fig:fig9} shows the spectrum for the disk component measured by the 
\fermilat{} (Table~\ref{table3}) 
and two model predictions (``naive'' and ``nominal'')  by \citet{Seckel1991} as described in Section~\ref{sec:disk}. 
In each set of curves, 
the lower bound (dotted line) is the CR induced \gray{} flux for the slant depth model,
and the upper
bound (solid line) is the \gray{} flux assuming showers are mirrored (as charged
particles would be). 
The observed spectrum can be well fitted by a single power-law with a spectral index of $2.11\pm0.73$.
The integral flux of the disk component 
is about a factor of 7 higher than predicted by the ``nominal'' model.
An obvious reason for the discrepancy could be the conditions of the unusually deep solar minimum during the reported observations.
However, this alone can not account for such a large factor, see a comparison with the EGRET data below.
Another possibility for an estimated ``nominal'' flux to be so low compared to the \fermilat{} observations 
is that the secondary particles produced by CR cascades 
exiting the atmospheric slab are ignored in the calculation while they are likely to reenter the Sun.
On the other hand, the proton spectrum by \citet{Webber1987} used in the calculation is about a factor of 1.5 higher above $\sim$6 GeV than that measured by the BESS experiment in 1998 \citep[see Figure 4 in][]{Sanuki2000}. 
Meanwhile, calculation of the disk emission relies on assumptions about CR transport in the inner heliosphere and in the immediate vicinity of the Sun thus allowing for a broad range of models (cf.\ ``naive'' vs.\ ``nominal'' models). 
The accurate measurements of the disk spectrum by the \fermilat{} thus warrant a new evaluation of the CR cascade development in the solar atmosphere.

The spectral shape of the observed disk spectrum is close to the predictions except below $\sim$230 MeV where the predicted spectral flattening is not confirmed by the observations. This may be due to the broad PSF making it difficult to distinguish between the components of the emission or a larger systematic error below $\sim$200 MeV associated with the IRFs.

The results of \fermilat{} observations can be also compared with
those from
the analysis of the EGRET data \citep{Orlando2008}. The latter gives an integral
flux ($\ge$100 MeV) for the disk component of $(1.8 \pm 1.1) \times 10^{-7}$ cm$^{-2}$ s$^{-1}$; for the IC
flux \citet{Orlando2008} obtain $(3.9 \pm 2.2) \times 10^{-7}$ cm$^{-2}$ s$^{-1}$ for elongation angles $\le$$10^\circ$.
The EGRET-derived integral IC flux is very close to the flux observed by the \fermilat{} for the same range of elongation angles
(Table~\ref{table1}).
Meanwhile, the integral flux of the disk component observed by the \fermilat{} (Table~\ref{table3}) is 
a factor of
$\sim$2.5 higher than obtained from EGRET data.
Such an increase in the \gray{} flux from the solar disk may imply a significant variation of the disk emission over the solar cycle. This is not surprising since the disk flux depends on the ambient CR 
proton spectrum in the immediate proximity of the solar atmosphere, in contrast to the solar IC emission, which is produced by electrons in a considerable part of the heliosphere and integrated along the line of sight. However, such variations should be confirmed by future observations.

The analysis reported here refers to the period of the solar minimum.
However, the solar activity
is just starting to pick up for the current cycle. We expect that the effects of the increased solar activity on the fluxes of the 
two components of the emission will be different and it is important to follow their evolution during the whole solar cycle. 
Continuous monitoring of the solar emission by the \fermilat{} over the whole 
solar cycle will enable us to study CR transport in the inner heliosphere, to improve models of the solar modulation, 
and of the development of CR cascades in the solar atmosphere. 

\acknowledgments

The \fermilat{} Collaboration acknowledges generous ongoing support
from a number of agencies and institutes that have supported both the
development and the operation of the \fermilat{} as well as scientific data analysis.
These include the National Aeronautics and Space Administration and the
Department of Energy in the United States, the Commissariat \`a l'Energie Atomique
and the Centre National de la Recherche Scientifique/Institut National de Physique
Nucl\'eaire et de Physique des Particules in France, the Agenzia Spaziale Italiana
and the Istituto Nazionale di Fisica Nucleare in Italy, the Ministry of Education,
Culture, Sports, Science and Technology (MEXT), High Energy Accelerator Research
Organization (KEK) and Japan Aerospace Exploration Agency (JAXA) in Japan, and
the K.~A.~Wallenberg Foundation, the Swedish Research Council and the
Swedish National Space Board in Sweden.
Additional support for science analysis during the operations phase is gratefully
acknowledged from the Istituto Nazionale di Astrofisica in Italy and the Centre National d'\'Etudes Spatiales in France.
I.~V.~M.\ and E.~O.\ acknowledge support from NASA grant NNX10AD12G.

\bibliography{ms} 


\newpage

\begin{deluxetable}{lcc}
\tabletypesize{\footnotesize}
\tablecaption{
Summary of the event selection cuts
\label{table4}}
\tablecolumns{3}
\tablewidth{0pt}
\tablehead{       
 &  Photons, & Livetime, \\
Cumulative event selections &   \% &  \%}
\startdata
$\theta_{\rm ROI} \le 20^\circ$ & 100 & 100\\
$|b_\odot | \ge 30^\circ $ & 29.2 & 60.9 \\
$\theta_{\rm Moon} > 20^\circ$ & 26.2 & 54.8 \\
$\theta(F_{\rm 1FGL}>2\times10^{-7}\ {\rm cm}^{-2}\ {\rm s}^{-1}) > 20^\circ$ & 6.5 & 17.6

\enddata
\end{deluxetable}

\begin{deluxetable}{cccccc}
\tabletypesize{\footnotesize}
\tablecaption{
Differential spectrum of the IC component
\label{table2}}
\tablecolumns{5}
\tablewidth{0pt}
\tablehead{       
Energy range, & \multicolumn{2}{c}{Flux$\pm$stat$\pm$syst, cm$^{-2}$ s$^{-1}$ MeV$^{-1}$} & \multicolumn{2}{c}{($E^2$Flux)$\pm$stat$\pm$syst, $10^{-5}$ cm$^{-2}$ s$^{-1}$ MeV}\\
\cline{2-3} \cline{4-5}
 $10^3$ MeV  & $\theta\le5^\circ$ & $\theta\le20^\circ$ & $\theta\le5^\circ$ & $\theta\le20^\circ$}
\startdata
$0.1-0.3$  & $(4.4\pm0.9 \pm0.1)\times10^{-10}$ & $(2.1\pm0.5\pm0.3)\times10^{-9\ }$
& $1.3\pm 0.2\pm 0.03$ & $6.2\pm 1.4\pm 0.1$
\smallskip\\
$0.3-1.0$  & $(4.9\pm1.0 \pm0.1)\times10^{-11}$ & $(2.2\pm0.5\pm0.3)\times10^{-10} $
& $1.4\pm0.3\pm 0.03$ & $6.1 \pm 1.5 \pm 0.9$
\smallskip\\
$1.0-3.0$  & $(5.5\pm1.1 \pm0.1)\times10^{-12}$ & $(2.2\pm0.6\pm0.2)\times10^{-11}$
& $1.6\pm0.3\pm 0.03$ & $6.1 \pm 1.8 \pm  0.6$
\smallskip\\
$3.0-10.$  & $(6.2\pm1.2 \pm0.1)\times10^{-13}$ & $(2.2\pm0.8\pm0.2)\times10^{-12}$
& $1.9\pm0.4\pm 0.03$  & $6.1\pm 2.4 \pm 0.6$
\medskip \\

 & \multicolumn{2}{c}{Integral flux$\pm$stat$\pm$syst, cm$^{-2}$ s$^{-1}$}  \smallskip\\
$\ge$0.1 & $(1.4\pm0.2_{-0.4}^{+0.5})\times10^{-7\ }$ & $(6.8\pm0.7_{-0.4}^{+0.5})\times10^{-7\ }$ 
\enddata
\tablecomments{The $E^2$Flux has been evaluated at the geometric mean of each bin.}
\end{deluxetable}

\begin{deluxetable}{cccccc}
\tabletypesize{\footnotesize}
\tablecaption{
Integral flux ($\ge$ 100 MeV) of the IC component
\label{table1}}
\tablecolumns{6}
\tablewidth{0pt}
\tablehead{       
Elongation         &          &\multicolumn{4}{c}{Integral flux, $10^{-7}$ cm$^{-2}$ s$^{-1}$}\\
\cline{3-6}
angle &  TS  & Flux$\pm$stat$\pm$syst & Model 1 & Model 2 & Model 3 
}
\startdata
$0^\circ-5^\circ$                  &  93   &$1.4\pm0.2_{-0.4}^{+0.5}$   & 1.15  & 1.33 &  1.78 \smallskip\\
$5^\circ-11^\circ$                  & 94    & $2.5\pm0.3_{-0.4}^{+0.1}$ &1.84 &  1.97 &  2.29\smallskip\\
$11^\circ-20^\circ$                &  43    & $3.0\pm0.5_{-0.2}^{+0.2}$  &3.11 &  3.19 &  3.50 \medskip\\

$0^\circ-10^\circ$  & 215 & $3.7\pm0.4_{-0.4}^{+0.5}$ &2.67 &  2.95 &  3.69 \smallskip\\
$0^\circ-20^\circ$  & 259 & $6.8\pm0.7_{-0.4}^{+0.5}$ & 6.10 &  6.48 &  7.57
\enddata
\end{deluxetable}

\begin{deluxetable}{cccc}
\tabletypesize{\footnotesize}
\tablecaption{
Differential spectrum of the disk component
\label{table3}}
\tablecolumns{4}
\tablewidth{0pt}
\tablehead{       
  Energy range,   &   &  Flux$\pm$stat$\pm$syst, &   ($E^2$Flux)$\pm$stat$\pm$syst, \\
  $10^3$ MeV  & TS  &  cm$^{-2}$ s$^{-1}$ MeV$^{-1}$ & $10^{-5}$ cm$^{-2}$ s$^{-1}$ MeV}
\startdata
$0.10-0.14$ & \ 59 & $(3.1 \pm 0.5 \pm0.6)\times10^{-9\ }$  & $4.3 \pm 0.7 \pm 0.9$\\
$0.14-0.19$ & \ 68 & $(2.4 \pm 0.3\pm 0.5)\times10^{-9\ }$ & $6.1 \pm 0.8 \pm 1.3$\\
$0.19-0.27$ & \ 85 & $(7.9 \pm 1.1 \pm 1.6)\times10^{-10}$  & $4.2 \pm 0.6 \pm 0.8$\\
$0.27-0.37$ & 100 & $(3.2 \pm 0.4 \pm 0.6)\times10^{-10} $ & $3.3 \pm 0.4 \pm 0.6$\\
$0.37-0.52$ & 222 &  $(2.2 \pm 0.2 \pm 0.4)\times10^{-10}$ & $4.2 \pm 0.4 \pm 0.8$\\
$0.52-0.72$ & 288 & $(1.2 \pm 0.1\pm 0.2)\times10^{-10}$ & $4.5 \pm 0.4 \pm 0.7$\\
$0.72-1.00$ & 243 & $(5.2 \pm 0.6 \pm 1.0)\times10^{-11}$  & $3.8 \pm 0.4 \pm 0.7$\\
$1.00-1.39$ & 282 & $(2.8 \pm 0.3\pm 0.6)\times10^{-11}$ & $3.9 \pm 0.4 \pm 0.8$\\
$1.39-1.93$ & 295 & $(1.6 \pm 0.2\pm 0.3)\times10^{-11}$ & $4.3 \pm 0.5 \pm 0.8$\\
$1.93-2.68$ & 193 & $(7.5 \pm 1.1 \pm 1.5)\times10^{-12}$ & $3.9 \pm 0.6 \pm 0.8$\\
$2.68-3.73$ & 167 & $(3.4 \pm 0.6 \pm 0.7)\times10^{-12}$ & $3.4 \pm 0.6 \pm 0.7$\\
$3.73-5.18$ & 143 & $(2.0 \pm 0.4 \pm 0.4)\times10^{-12}$ & $3.9 \pm 0.8 \pm 0.8$\\
$5.18-7.20$ & \ 85 & $(8.2 \pm 2.1 \pm 1.6)\times10^{-13}$ & $3.1 \pm 0.8 \pm 0.6$\\
$7.20-10.0$ & \ 21& $(2.2 \pm 0.9 \pm 0.4)\times10^{-13}$ & $1.6 \pm 0.6 \pm 0.3$  
\medskip\\

&  \multicolumn{2}{r}{Integral flux$\pm$stat$\pm$syst, cm$^{-2}$ s$^{-1}$}  \smallskip\\
$\ge$0.1  & & $(4.6 \pm 0.2^{+1.0}_{-0.8})\times10^{-7\ }$
\enddata
\tablecomments{The $E^2$Flux has been evaluated at the geometric mean of each bin.}
\end{deluxetable}

\clearpage

\end{document}